\documentclass[a4paper,11pt]{article}
\usepackage{pos}

\title{Axion-like particle constraints from preSN in future experiments}

\author*[a]{Federica Giacchino}

\affiliation[a]{Istituto Nazionale di Fisica Nucleare - Sezione Roma Tor Vergata,\\
  via della Ricerca Scientifica 1, 00133 Rome, Italty}

\emailAdd{federica.giacchino@roma2.infn.it}

\abstract{We extend the study of the search for ALP-photon emissions from a Supernovae (SNe) progenitor, the Red Supergiant Star (RGS), as Betelgeuse, arising through a combination of Bremsstrahlung, Compton, and Primakoff processes, in the hard-X and MeV energy range, using the next proposed and future experiments, such as COMCUBE, GECCO, COSI, AMEGO-X, and e-ASTROGAM.}

\FullConference{2nd Training School and General Meeting of the COST Action COSMIC WISPers  (CA21106) (COSMICWISPers2024)\\
 10-14 June 2024 and 3-6 September 2024\\
Ljubljana (Slovenia) and Istanbul (Turkey)\\}

\begin{document}
\maketitle
\section{Introduction}
\label{sec:intro}
Axion-like particles (ALPs) are pseudo-Nambu-Goldstone bosons that arise from the explicit breaking of an approximate shift symmetry. They are predicted by several Beyond Standard Model (BSM) theories and are consequently motivated by various extensions of the Standard Model (SM), such as the QCD axion~\cite{Peccei:1977hh}. A key distinguishing feature of ALPs is that their couplings and mass, $m_a$, are independent parameters, allowing the mass to span a broad range of values. ALPs could serve as dark matter candidates for $m_a<$ keV or as \emph{portals} to a hidden sector at the MeV scale. Recently, Feebly Interacting Particles (FIPs) have garnered attention as compelling alternatives to the widely considered dark matter candidate, the Weakly Interacting Massive Particle (WIMP). These light particles, well below the weak scale, are theorized to reside in a hidden sector, where dark matter could represent only a glimpse of more fundamental ultraviolet-complete theories. The ALPs are FIPs promising candidate. Over the years, many experimental approaches, spanning astrophysical to accelerator-based frameworks, have been dedicated to exploring these possibilities (see for example these recent reviews~\cite{DiLuzio:2020wdo},~\cite{Caputo:2024oqc}).
\section{ALP from the space}
\label{sec:ALP}
ALPs can be produced in various astrophysical sources through different mechanisms. Once generated, these particles travel through intergalactic space, and when they interact with external magnetic fields, they generate a distinctive photon signal via ALP-photon conversion, which could be detectable by space-based experiments. In recent years, significant research has focused on the detection of ALPs. Here, I concentrate on a specific study of an astrophysical source within a promising yet underexplored energy range.

Notably, the Lagrangian of the ALP with SM field interaction is described by $
\mathcal{L}_{ALP}=\frac{1}{2}(\partial^{\mu} a)^2-\frac{1}{2}m^2_aa^2 - \frac{1}{4}g_{a\gamma}aF_{\mu\nu}\tilde{F}^{\mu\nu}-i\sum_f g_{af}\,m_f\,a\,\bar{\psi}_f\,\gamma_5\,\psi_f $. Stars are efficient factories of FIPs, such as ALPs in our case, producing them in large quantities. ALPs are generated efficiently in stellar cores and stream freely due to their feeble interactions. As described in~\cite{Raffelt:2006cw}, the dominant production mechanism varies with the evolutionary stage of the star, allowing tests of specific ALP couplings. The Primakoff process, $\gamma + Z_e \rightarrow Z_e + a$, which probes the ALP-photon coupling $g_{a\gamma}$, is particularly relevant for main-sequence stars like the Sun and for horizontal branch (HB) stars. The ALP-electron interaction, tested through the Compton process $\gamma + e \rightarrow e + a$, also contributes to HB stars. Additionally, the Bremsstrahlung process, $e + Z_e \rightarrow e + Z_e + a$, is significant in red giant branch (RGB) stars and white dwarfs. For neutron stars and SNe processes which involves interactions with nucleon are dominants. There are around $20$ supergiants stars within $1$ kpc from the Sun. Some of them in advanced stage which could produce a large ALP flux. One of them is Alpha Orionis, aka Betelgeuse. It is a red supergiants located at $d\sim 200$ pc, with spectral type M2Iab, and mass $\sim 15 - 24$ M$_{\odot}$. It is a good target to search for ALP-induced photon emission using future astronomical instruments. As said before, the most efficient axion production mechanisms in the core of such massive star in an advanced evolutionary stage are the Primakoff conversion, Compton scattering and Bremsstrahlung (ALP-nucleon coupling is subdominant):
\begin{eqnarray}
\frac{d\dot{N}_a}{dE} &=& C^P \Big(\frac{g^2_{a\gamma}}{10^{-11}}\Big)\,\text{GeV}^{-1}\Big(\frac{E}{E^P_0}\Big)^{\beta^P}e^{-(\beta^P+1)E/E^P_0} \nonumber\\
& & +\,C^C \Big(\frac{g^2_{ae}}{10^{-13}}\Big)\,\text{GeV}^{-1}\Big(\frac{E}{E^C_0}\Big)^{\beta^C}e^{-(\beta^C+1)E/E^C_0} \nonumber \\
& &+\,C^B \Big(\frac{g^2_{ae}}{10^{-13}}\Big)\,\text{GeV}^{-1}\Big(\frac{E}{E^B_0}\Big)^{\beta^B}e^{-(\beta^B+1)E/E^B_0}
\label{eq:ALPflux}
\end{eqnarray}
Observations of Betelgeuse's surface indicate that its luminosity and temperature remain constant, leaving its precise evolutionary stage uncertain. To model a possible stage in Betelgeuse's evolution, stellar profiles were computed using the Full Network Stellar Evolution code (FuNS)~\cite{Straniero:2019dtm}. In~\cite{Xiao:2022rxk}, Table 1 presents $12$ numerical models with corresponding parameter values from Eq.~\ref{eq:ALPflux}, from the less evolved state in \emph{model 0} to the Ne-burning phase in \emph{model 11}.  Once produced, an ALP beam of energy $E$ propagating along $z-$axis in the presence of an external magnetic field $\vec{B}$ can convert due to the mixing matrix of ALP-photon interaction. In the approximation regime, $E >> m$ and transverse magnetic field $B_T$ homogeneous, the ALP-photon conversion probability is 
\begin{equation}
P_{a\gamma}=8.7\times10^{-6}\Big(\frac{g_{a\gamma}}{10^{-11}}\text{GeV}^{-1}\Big)^2\Big(\frac{B_T}{1\mu\text{G}}\Big)^2 \Big(\frac{d}{197\,\text{pc}}\Big)^2 \frac{\sin^2 (qd)}{(qd)^2}
\label{eq:conversionprobability}
\end{equation}
with $d$ the distance travels and $qd$ the product of the momentum transfer and the magnetic field length which depends on ALP mass, electron density and photon energy. Here $qd<<1$ ensuring that the conversion probability is energy independent and the photon spectrum keeps the same shape of the original ALP distribution. After converted, we can detect with space-based experiments the photon produced $\frac{dN_{\gamma}}{dE\,dS\,dt} = \frac{1}{4\pi d^2}\frac{d\dot{N}_a}{dE}P_{a\gamma}$. A dedicated NuSTAR observation has already searched for axions produced in the plasma inside Betelgeuse, leading to stringent bounds on the axion-photon coupling in the sub-neV mass region~\cite{Xiao:2022rxk}.  For the couplings of interest, the resulting photon flux can then be searched in a wide energy band, from hard X-ray to gamma-ray. Remarkably, NuSTAR is limited at high energies, with a sensitivity which rapidly drops above $80$ keV. Therefore we need to implement the analysis in various Compton telescopes designed to cover the so-called \emph{MeV gap}. This energy range is poorly explored by the current generation of instruments, and the proposed ``ASTROMEV'' space missions are essential, not only for this purpose but also for other astrophysical motivations. 
\section{ASTROMEV experiments: sensitvity and results}
\label{sec:experiments}
COMCUBE~\cite{Laviron:2021} is a proposed European Project, a CubeSat mission functioning as a Compton polarimeter, designed to detect gamma-ray emissions from astrophysical sources such as gamma-ray bursts. It is expected to operate in the energy range from $100$ keV to $1.5$ MeV, with an angular resolution of $\sim 20$ deg and an energy resolution $(3-13)\%$. One COMCUBE is composed by 4 Unit which the single unit is formed by a tracker, one layer of plastic scintillator and a calorimeter. We have included the nano-satellites for a cost-effective solution and the easily scaled in number, enabling a more economical implementation. GECCO is an American proposed mid-sized explorer class mission which optimizes for gamma-ray detection in the Compton regime (from 100 keV up to $\sim $10 MeV) through the use of a Compton telescope combined with a deployable coded aperture mask telescope for the photoelectric regime~\cite{2022JCAP...07..036O}. COSI~\cite{Tomsick:2019wvo}, a NASA approved medium-size mission, is set to launch in 2027. Its goals include studying positron annihilation, stellar nucleosynthesis, polarization of GRBs and AGNs, and the localisation of short-GRB for multi-messager astrophysics. COSI will detect gamma rays in the energy range between $200$ keV to $8$ MeV. AMEGO-X (All-sky Medium Energy Gamma-ray Observatory eXplorer) is a NASA-proposed mission consisting of two detector subsystems: a gamma-ray detector and an anti-coincidence detector~\cite{AMEGO:2019gny}. The mission will survey the sky across a wide energy range, from $100$ keV to $1$ GeV, with unprecedented sensitivity. AMEGO-X will detect gamma rays through both Compton interactions and pair production, bridging the sensitivity gap between hard X-rays and high-energy gamma rays. The European counterpart to AMEGO-X is e-ASTROGAM~\cite{e-ASTROGAM:2017pxr}, with a similar setup. It emphasizes the multi-messenger approach and the polarization of gamma-ray sources, focusing on cosmic phenomena. The sensitivities for all the considered experiments are reported in the Table~\ref{tab:sensitivity}.
\begin{table}[t!]
\begin{tabular}{c|c|c|c|c|c|c}
\hline
\hline
En & 1 C & 64 C & COSI & GECCO & e-ASTRO & AMEGO-X \\
\hline
$100$ & $7.9\times10^{-11}$ & $8.1\times10^{-12}$ & - & $1.1\times10^{-11}$ & - & $3\times 10^{-10}$ \\
$500$ & $2.4 \times 10^{-10}$ & $1.7\times10^{-11}$ & $3.8\times10^{-11}$ & $1.3\times10^{-11}$ & $1.9\times 10^{-12}$ & $8.7\times10^{-12} $ \\
$1000$ & $5.1 \times 10^{-10}$ & $3.1 \times10^{-11}$ & $5.8\times10^{-11}$ & $1.3\times10^{-11}$ & $3.4\times10^{-12}$ & $ 6.8\times 10^{-12}$ \\
$1500$ & $2 \times 10^{-9}$ & $9\times10^{-11}$ & $8.2\times 10^{-11}$ & $1.5\times10^{-11}$ & $3.6\times 10^{-12}$ & $ 6.5\times 10^{-12}$ \\
$5000$ & - & - & $3.9\times 10^{-10}$ & $3\times10^{-11}$  & $6.8\times 10^{-12}$ & $2\times 10^{-11}$ \\
$10000$ & - & - & - & - & $7.1\times10^{-12}$ & $1.5\times 10^{-11}$\\
\hline
\hline
\end{tabular}
\caption{Table of sensitivity in erg/cm$^2$/s for some Energies in keV of 1 COMCUBE~\cite{Laviron:2021}, 64 COMCUBE, GECCO~\cite{2022JCAP...07..036O}, COSI~\cite{Tomsick:2019wvo}, e-ASTROGAM~\cite{e-ASTROGAM:2017pxr}, and AMEGO-X~\cite{AMEGO:2019gny}.}
\label{tab:sensitivity}
\end{table}
The preliminary results of the sensitivity of ALP parameter space show a discovery potential with the ASTROMEV experiments. The transverse magnetic field $B_T=1.4\,\mu G$, $d=200$ pc and electronic density $n_e = 0.013$ are the benchmark values we have considered. The background for COMCUBE is simulated using the MegaLib tool, while for the other experiments, we have convolved FLUKA simulations of the expected flux in the direction of Betelgeuse with the effective area of each experiment. The instrumental response functions of the experiments are not included in the simulation, which leads to an underestimation of the background. Using $50\times 10^3$ seconds observation of Betelgeuse by the previous satellite telescopes, we set $95\%$ C.L. upper limits on the ALP-electron ($g_{ae}$) and ALP-photon ($g_{a\gamma}$) couplings. The \emph{model 11} is the most promising stellar model for the future experiments. Notably, for mass $m_a=3.5\times 10^{-11}$ eV, we find a discovery potential region in $g_{ae} < 2 \times 10^{-13} $ for $g_{a\gamma} < 8 \times 10^{-11}$ GeV$^{-1}$ for e-ASTROGAM, COSI, GECCO and COSI. For masses $m_a < 10^{-10}$ eV, we find $ (1 < g_{a\gamma}\times g_{ae} <  4) \times 10^{-25} $ GeV $^{-1}$ for GECCO and e-ASTROGAM,  $ (3 < g_{a\gamma}\times g_{ae} <  4)\times 10^{-25} $ GeV $^{-1}$ for COSI, while for $m_a > 10^{-10}$ eV also AMEGO-X and 64 COMCUBE cover some regions of couplings not yet explored by current experiments.
\section{Conclusion}\label{sec:conclusions}
In this work we want to support next generation of future experiments with a new physics scientific topic, the photon signal of ALP origin from supergiants as Betelgeuse. The preliminary results show a discovery potential in the ALP parameter space.
\section{Acknowledgments}
This article is based on the work from COST Action COSMIC WISPers CA21106, supported
by COST (European Cooperation in Science and Technology).

\end{document}